\documentclass[longauth]{aa}

\usepackage{graphicx}
\usepackage{hyperref}
\usepackage{natbib}                 
\bibpunct{}{}{;}{a}{}{,}
\usepackage{amsmath}                
\usepackage{txfonts}

\begin{document}
   \title{Discovery and follow-up studies of the extended, off-plane, VHE gamma-ray source HESS J1507-622}


{\small
\author{H.E.S.S. Collaboration
 \and F.~Acero \inst{15}
 \and F. Aharonian\inst{1,13}
 \and A.G.~Akhperjanian \inst{2}
 \and G.~Anton \inst{16}
 \and U.~Barres de Almeida \inst{8} \thanks{supported by CAPES Foundation, Ministry of Education of Brazil}
 \and A.R.~Bazer-Bachi \inst{3}
 \and Y.~Becherini \inst{12}
 \and B.~Behera \inst{14}
 \and K.~Bernl\"ohr \inst{1,5}
% ...
}
}

\newpage

\institute{\small
Max-Planck-Institut f\"ur Kernphysik, P.O. Box 103980, D 69029
Heidelberg, Germany
\and
 Yerevan Physics Institute, 2 Alikhanian Brothers St., 375036 Yerevan,
Armenia
\and
Centre d'Etude Spatiale des Rayonnements, CNRS/UPS, 9 av. du Colonel Roche, BP
4346, F-31029 Toulouse Cedex 4, France
\and
Universit\"at Hamburg, Institut f\"ur Experimentalphysik, Luruper Chaussee
149, D 22761 Hamburg, Germany
\and
Institut f\"ur Physik, Humboldt-Universit\"at zu Berlin, Newtonstr. 15,
D 12489 Berlin, Germany
\and
LUTH, Observatoire de Paris, CNRS, Universit\'e Paris Diderot, 5 Place Jules Janssen, 92190 Meudon, 
France
\and
IRFU/DSM/CEA, CE Saclay, F-91191
Gif-sur-Yvette, Cedex, France
\and
University of Durham, Department of Physics, South Road, Durham DH1 3LE,
U.K.
\and
Unit for Space Physics, North-West University, Potchefstroom 2520,
    South Africa
\and
Laboratoire Leprince-Ringuet, Ecole Polytechnique, CNRS/IN2P3,
 F-91128 Palaiseau, France
\and 
Laboratoire d'Annecy-le-Vieux de Physique des Particules,
Universit\'{e} de Savoie, CNRS/IN2P3, F-74941 Annecy-le-Vieux,
France
\and
Astroparticule et Cosmologie (APC), CNRS, Universite Paris 7 Denis Diderot,
10, rue Alice Domon et Leonie Duquet, F-75205 Paris Cedex 13, France
\thanks{UMR 7164 (CNRS, Universit\'e Paris VII, CEA, Observatoire de Paris)}
\and
Dublin Institute for Advanced Studies, 5 Merrion Square, Dublin 2,
Ireland
\and
Landessternwarte, Universit\"at Heidelberg, K\"onigstuhl, D 69117 Heidelberg, Germany
\and
Laboratoire de Physique Th\'eorique et Astroparticules, 
Universit\'e Montpellier 2, CNRS/IN2P3, CC 70, Place Eug\`ene Bataillon, F-34095
Montpellier Cedex 5, France
\and
Universit\"at Erlangen-N\"urnberg, Physikalisches Institut, Erwin-Rommel-Str. 1,
D 91058 Erlangen, Germany
% ...
}

\offprints{O.~Tibolla\\
  \email{omar.tibolla@astro.uni-wuerzburg.de}}

   \date{Received 10 June 2010; accepted 7 October 2010}

 
  \abstract
   {The detection of gamma-rays in the very-high-energy (VHE) range (100 GeV-100 TeV) offers the possibility of studying the parent population of ultrarelativistic particles found in astrophysical sources, so it is useful for understanding the underlying astrophysical processes in nonthermal sources.}
   {The discovery of the VHE gamma-ray source HESS J1507-622 is reported and possibilities regarding its nature are investigated.}
   {The H.E.S.S. array of imaging atmospheric Cherenkov telescopes (IACTs) has a high sensitivity compared with previous instruments ($\sim$1\% of the Crab flux in 25 hours observation time for a 5$\sigma$ point-source detection) and has a large field of view (${\sim}5^{\circ}$ in diameter).
HESS J1507-622 was discovered within the ongoing H.E.S.S. survey of the inner Galaxy, and the source was also studied by means of dedicated multiwavelength observations. }
   {A Galactic gamma-ray source, HESS J1507-622, located ${\sim}3.5^{\circ}$ from the Galactic plane was detected with a statistical significance $>$9 $\sigma$. Its energy spectrum is well fitted by a power law with spectral index $\Gamma = 2.24 \pm 0.16_{stat} \pm 0.20_{sys}$ and a flux above 1 TeV of $(1.5 \pm 0.4_{stat} \pm 0.3_{sys}) \times 10^{-12}$ cm$^{-2}$ s$^{-1}$. Possible interpretations ( considering both hadronic and leptonic models) of the VHE gamma-ray emission are discussed in the absence of an obvious counterpart.}
   {}

   \authorrunning{H.E.S.S. Collaboration}
   \titlerunning{HESS J1507-622}
   \maketitle
%

\section{Introduction}


In the extension of the successful H.E.S.S. survey of the Galactic plane (\cite{survey}, \cite{survey2}), performed in 2006/2007 with the High Energy Stereoscopic System (H.E.S.S.), a number of new sources have been discovered and many of them are still unidentified (\cite{unid}).
HESS J1507-622 is among the brightest ($\sim$8\% of Crab Flux) newly discovered sources, and it lacks plausible counterparts, as is the case for HESS J1427-608, HESS J1708-410 (\cite{unid}), and HESS J1616-508 (\cite{survey2}). While all unidentified VHE sources that have been discovered in the H.E.S.S. Galactic plane scan are so far located within $\pm$1 degree from the Galactic equator, HESS J1507-622 is unique in this respect since it lies $\sim$3.5$^\circ$ from the Galactic plane. Considering the comparably low hydrogen column density $n_H$ at 3.5$^{\circ}$ off the plane, hence the lower Galactic absorption in X-rays and the reduced background emission, one would expect to detect a bright counterpart despite the anticipated lower spatial source density of Galactic counterparts.

Most Galactic VHE emitters are connected to young stellar populations, usually concentrated near the Galactic disk. Therefore, it is quite surprising to find an unidentified VHE gamma-ray source with a 3.5$^\circ$ offset from the Galactic plane, unless the source is nearby. HESS J1507-622 may offer the intriguing possibility of constraining the environment of an object that emits very-high energy radiation but lacks obvious counterparts at lower energies. Our line of sight towards the source direction intersects the Galactic disk with a scale height of 50 pc up to a distance of, at most, 1 kpc extending farther into the Galactic halo. 

%





\section{Observations}

\subsection{H.E.S.S. observations and data analysis}

Within the 2006/2007 extension of the H.E.S.S. survey of the Galactic plane there were indications of a source at the position of HESS J1507-622; follow-up dedicated observations were performed for 9.7 hours (21 runs of $\sim$28 minutes each) in 2007 and in 2008. The data are calibrated as detailed in Aharonian et al. (2004). A thorough discussion of the H.E.S.S. standard analysis and performance of the instrument can be found in Aharonian et al. (2006c).


   \begin{figure}
   \centering
   \includegraphics[width=8cm,height=7.3cm]{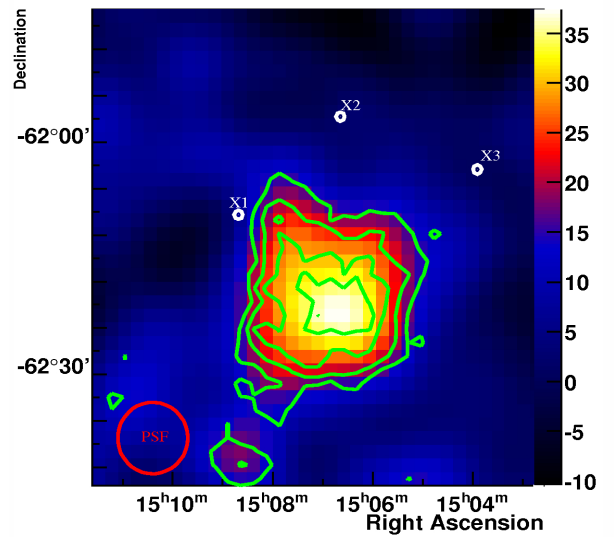}
      \caption{Excess count map (smoothed with a Gaussian of $0.07^{\circ}$
radius) of HESS J1507-622. The green contours show 3$\sigma$, 4$\sigma$, 5$\sigma$,
6$\sigma$ contour levels for an integration radius of $0.12^{\circ}$. The white circles represent the position of three faint RASS (\cite{faintRASS}) sources: X1 indicates 1RXS J150841.2-621006, X2 indicates 1RXS J150639.1-615704 and X3 indicates 1RXS J150354.7-620408. The 68\% containment radius for the point spread function (PSF) of the H.E.S.S. instrument for these observations is shown in magenta.
              }
         \label{1507}
   \end{figure}

The peak significance, calculated following the method of Li and Ma (1983), is 9.3$\sigma$ for the 9.7 hours of dedicated observations (using a $0.22^{\circ}$ oversampling radius, which is the standard radius used in source searches in the H.E.S.S. Galactic plane survey).
Figure \ref{1507} shows the uncorrelated excess count map (smoothed with Gaussian of $0.07^{\circ}$) for the dedicated observations, using \emph{hard cuts} (\cite{crab}) and the \emph{ring background method} (\cite{berge}).
A two-dimensional Gaussian fit yields the best position at RA = $226.72^{\circ} \pm 0.05^{\circ}$, Dec = $-62.35^{\circ} \pm 0.03^{\circ}$, and the source is slightly extended with intrinsic size (not including the PSF) of $0.15^{\circ} \pm 0.02^{\circ}$ radius.
The peak significance for this smaller ($0.12^{\circ}$, chosen to reveal the source morphology at a reasonable significance without oversmoothing) correlation radius is 7.0$\sigma$.

   \begin{figure}
   \centering
   \includegraphics[width=9cm]{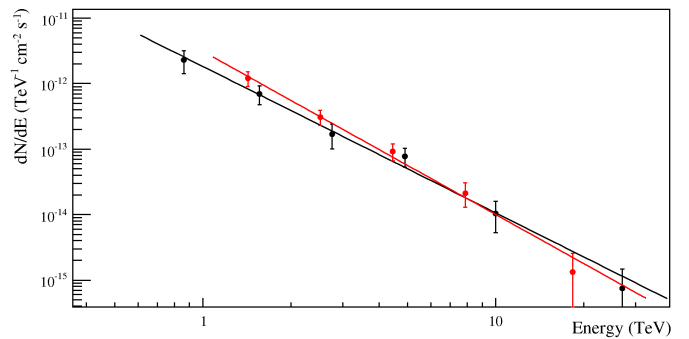}
      \caption{The VHE spectrum observed from HESS J1507-622. The black line represents the best $\chi^2$ fit of a power law to observed data using the \emph{standard cuts}, whereas the red line shows the spectral result using \emph{hard cuts} \citep[as described in][]{crab}.
              }
         \label{spectra}
   \end{figure}

The energy spectrum is reconstructed ($0.22^{\circ}$ extraction radius, which is larger than the source, since the morphology of the source is not well known) with the method \citep[presented in][]{crab} with background subtracted using the \emph{reflected-region method} (\cite{berge}): the results are shown in Figure \ref{spectra}.
Using \emph{standard cuts} (\cite{crab}), hence a lower energy threshold ($\sim~500$~GeV), the observed spectrum can be well fitted with a power-law $dN/dE = k (E/1$ TeV$)^{-\Gamma}$ with photon index $\Gamma = 2.24 \pm 0.16_{stat} \pm 0.20_{sys}$ and a flux normalization $k = (1.8 \pm 0.4) \times 10^{-12}$ TeV$^{-1}$ cm$^{-2}$ s$^{-1}$, making the integral flux (above 1 TeV) $(1.5 \pm 0.4_{stat} \pm 0.3_{sys}) \times 10^{-12}$ cm$^{-2}$ s$^{-1}$.
Using \emph{hard cuts} (energy threshold $\sim~1$~TeV) and hence a better gamma-hadron separation (\cite{crab}), the observed spectrum is well fitted by a power-law with photon index $\Gamma = 2.49 \pm 0.18_{stat} \pm 0.20_{sys}$ and a flux normalization $k = (3.1 \pm 0.8) \times 10^{-12}$ TeV$^{-1}$ cm$^{-2}$ s$^{-1}$, yielding an integral flux (above 1 TeV) of $(2.1 \pm 0.6_{stat} \pm 0.4_{sys}) \times 10^{-12}$ cm$^{-2}$ s$^{-1}$. The data points are compatible and the difference in flux arises from slope difference and from extrapolation to 1 TeV. 
No significant evidence of curvature in the spectrum has been found. An independent analysis and calibration based on a fit of camera images to a shower model (\cite{model2D}) gives compatible results.







%


%


\subsection{X-ray observations and data analysis}

The source was observed with XMM-Newton on Jan 27, 2009 for 28 ks and with Chandra on Jun 6, 2009 for 20 ks.

\begin{figure}
   \centering
   \includegraphics[width=9cm]{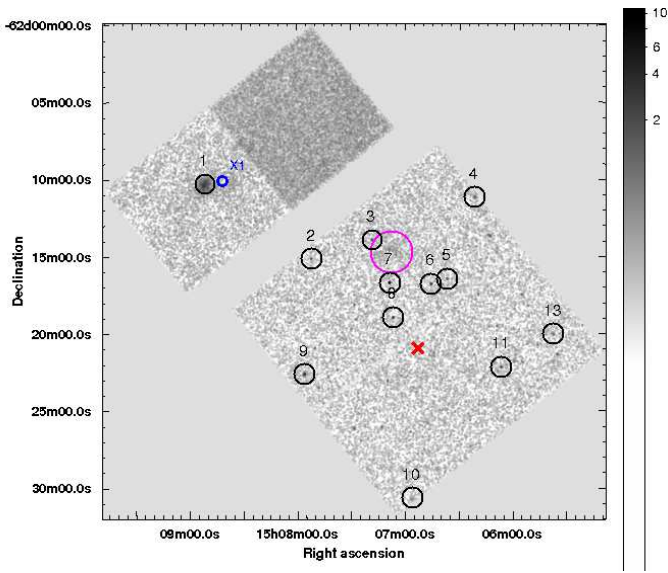}
      \caption{Smoothed and background-subtracted count map of Chandra observation of HESS J1507-622. Black circles indicate the 12 sources detected by {\tt celldetect}. The blue circle indicates 1RXS J150841.2-621006, centered on its nominal position and with radius corresponding to its positional uncertainities (\cite{faintRASS}). The faint extended emission described in the text is shown in magenta. The red cross indicates the centroid position of HESS J1507-622.
              }
         \label{chandra}
   \end{figure}

Unfortunately, the occurrence of a huge soft proton flare dramatically affected the 28 ks observation with XMM-Newton, leading to a good time interval (GTI) of only 0.8 ks for the PN detector, 8.0 ks for MOS1 and 9.2 ks for MOS2 detector. The soft proton flare is identified by extracting the high-energy ($>$ 10 keV) light curve of the whole observation and is confirmed by the radiation monitor onboard XMM-Newton. However, background proton events still affect the remaining GTI, as they lead to wrong vignetting correction, in particular in the outer regions of the MOS detectors. The data were reprocessed and the GTI was analyzed with SAS 9.0, discovering one point-like source (source 7 in Fig. \ref{chandra}) with a flux of $(1.4 \pm  0.4)  \times 10^{-13} $erg cm$^{-2}$s$^{-1}$ in the energy range 2-10 keV: a compatible flux value has been found within Chandra observations and the point-like source is identified as a star (Tyc 9024.1705.1).

Fig. \ref{chandra} shows the smoothed and background subtracted count map between 0.3 and 8 keV as seen by Chandra for 20 ks. 
Chandra data were analyzed using CIAO 4.1. The background was estimated using the standard blank sky field files, since HESS J1507-622 covers the whole ACIS-I field of view. Eleven point-like sources (mainly identified as stars) and one extended source have been discovered using the source detection algorithm {\tt celldetect}.

One of the most interesting discoveries by this Chandra observation concerns 1RXS J150841.2-621006 (source 1 in Fig. \ref{chandra}). This source is extended (35-40 arcsec) and spatially coincident with the radio source MGPS J150850-621025 (\cite{molonglo}). The source is detected with an energy flux of $(7.0 \pm  0.7)  \times 10^{-13} $erg cm$^{-2}$s$^{-1}$  in the range 2-10 keV, compatible with the flux measured by ROSAT. Even if the position of 1RXS J150841.2-621006 (\cite{faintRASS}) does not coincide with the position of the extended Chandra source, the two positions are compatible when considering the well known systematics in the positional accuracy of ROSAT (\cite{ROSATsys}).
The nature of the extended source 1RXS J150841.2-621006 is still unclear, since its spectrum is well fitted ($\chi^2$/d.o.f. = 30.4/36) by a simple power law (spectral index $= 2.0 \pm 0.3$; $n_H = (8.2 \pm 2.7) \times 10^{21}$ cm$^{-2}$), but the fit is equally good ($\chi^2$/d.o.f. = 34.7/36) for a thermal plasma scenario (kT $= 5^{+ 5}_{- 2}$ keV; $n_H = (7.0 \pm 2.7) \times 10^{21}$ cm$^{-2}$).
However, given its offset from the VHE source emission peak, outside the 3$\sigma$ significance contour, 1RXS J150841.2-621006 is not a plausible counterpart for our source.

The other important aspect of the Chandra analysis is the existence of a faint extended source that escaped the source detection algorithm {\tt celldetect} but is detected (116 counts over a level of background of 32 counts) by {\tt vtpdetect}, which performs better for faint diffuse structures. This source seems extended (20-25 arcsec), and it is slightly above the background level. Its flux is  $(1.1^{+0.3}_{-0.5})  \times 10^{-13} $erg cm$^{-2}$s$^{-1}$ in the energy band between 2 and 10 keV, resulting from a power-law fit to the spectrum extracted, using {\tt specextract}, from a region of 50 arcsec radius, centered on RA$_{J2000} = 15^{\rm{h}} 07^{\rm{m}} 06.8^{\rm{s}}$ , dec$_{J2000} = -62^{\circ} 14' 45.0''$.
This faint source is confirmed by means of {\tt wavdetect} with a significance of 6.9$\sigma$. 
It has a much smaller angular size than the VHE source, so an association is not obvious with current data, although the possibility remains that it is a bright part of a larger, weaker source.  There is unfortunately no strong conclusion that can be made, but it could be linked to an old pulsar wind nebula (PWN) still visible at VHE. In fact the VHE PWNe sizes generally increase with pulsar age while the X-ray PWNe sizes show the opposite trend; moreover, for pulsars older than $\sim 10^3$ years the VHE PWNe are typically 100-1000 times larger than the sizes of the X-ray PWNe (while the difference is only a factor 2 for some younger PWNe, like the Crab Nebula), as shown by \cite{kar} (e.g. HESS J1718-385). However, in the case of HESS J1507-622, there is no pulsar emission detected, as discussed in the following section.


\section{Discussion}

The unique feature that distinguishes HESS J1507-622 from any other unidentified VHE source is its angular offset ($\sim 3.5^{\circ}$) from the Galactic plane. Such an offset may mean that HESS J1507-622 is a nearby, local object or that it is truly distant in the halo. This second possibility may have important implications for any gamma-ray production scenario. At a location away from the Galactic disk, the density of target material is quite low (\cite{lockman84}), making a hadronic scenario for HESS J1507-622 less favorable. On the other hand, the cosmic microwave background (CMB), as target photon field for gamma-ray production, is available everywhere pointing towards a leptonic gamma-ray emission mechanism. However, if confirmed by deeper observations, the lack of emission at other wavebands could imply a hadronic origin for HESS J1507-622. 
Both cases are discussed in this section.

\subsection{Searching for counterparts}

\begin{figure}
   \centering
   \includegraphics[width=9cm]{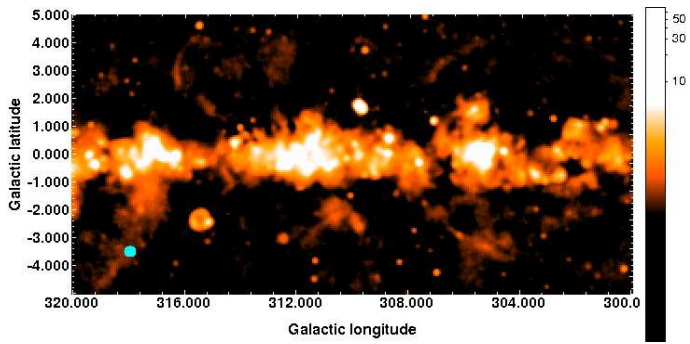}
      \caption{2.4 GHz radio map from \cite{duncan}. HESS J1507-622 is marked by a cyan circle centered at the nominal position of the VHE gamma-ray source and with 0.15$^{\circ}$ radius.
              }
         \label{24ghz}
   \end{figure}

\begin{figure}
   \centering
   \includegraphics[width=9cm]{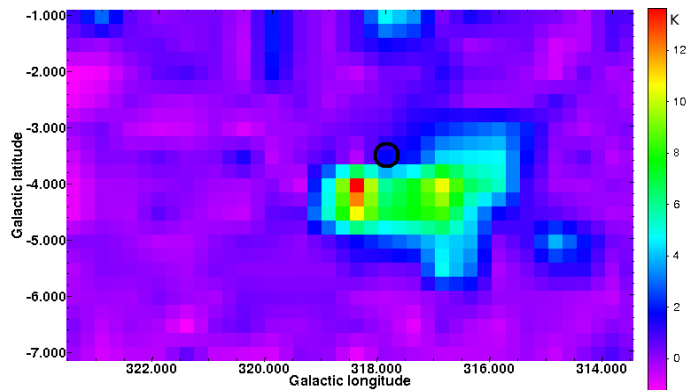}
      \caption{CO map of the region around HESS J1507-622 integrated in the velocity parameter in the range -10 to 0 km/s; the CO peak is at $\sim$-5 km/s, which corresponds to a distance of $\sim$400 pc (or, more unlikely 12 kpc). The black circle represents HESS J1507-622: it is centered at the nominal position of the VHE gamma-ray source and has a 0.15$^{\circ}$ radius. 
              }
         \label{damecat}
   \end{figure}

The search for counterparts was made following the procedure outlined in Aharonian et al. (2008).
At radio and infrared wavelengths, HESS J1507-622 is offset too far from the Galactic plane to be covered by the Southern Galactic Plane Survey (\cite{haverkorn}) or by Spitzer GLIMPSE (\cite{Spitzer}). This region of the sky had been covered by the Midcourse Space Experiment (MSX) in all its wave bands (8.28 $\mu$m, 12.13 $\mu$m, 14.65 $\mu$m, 21.3 $\mu$m, \cite{msx}) and by the MOLONGLO Galactic plane survey (\cite{molonglo}), yielding no evidence of any plausible counterpart. HESS J1507-622 is located on a radio emission filament shown in \cite{duncan} at 2.4 GHz, which was tentatively considered a part of a very large ($\sim 15^{\circ}$ in diameter) and nearby candidate SNR (visible in Fig. \ref{24ghz}). In the complete CO survey (\cite{Dame}), the H.E.S.S. source lies near the edge of a large ($\sim 5^{\circ} \times \sim 2^{\circ}$) nearby CO molecular cloud (visible in Fig. \ref{damecat}), and the peak velocity of this cloud, around -5 km/s, would most likely place it quite near at a distance of $\sim$400 pc. The substantial difference in extension and, in the case of the CO molecular cloud, the offset of $\sim 1^{\circ}$from the HESS source centroid, suggest no obvious scenario for an association.

The absence of obvious counterparts in X-rays in combination with the observed VHE flux can be used to compare the multiwavelength properties of HESS J1507-622 with other unidentified sources. The ratio of F$_{\gamma}$(1-10 TeV)/F$_\mathrm{X}$(2-10 keV) has been suggested as a viable parameter for such a comparison (Yamazaki et al. \cite{yamazaki06}). If the flux of the Chandra faint extended source (\ref{chandra}) is adopted as an upper limit for any low-energy counterpart and using the H.E.S.S. flux found with hard cuts for HESS J1507-622, this ratio is estimated to be $>$100. This would place the source above the ratio of F$_{\gamma}$/F$_\mathrm{X}$ for the first unidentified source of the H.E.S.S. Galactic plane scan HESS J1303-631, which was estimated to be $>$1.6. Moreover, this ratio for HESS J1507-622 would be greater than the value obtained for the ``darkest'' VHE source so far of $>$55 for HESS J1616-508 (Matsumoto et al. \cite{matsumoto07}).  The caveat of determining the ratio F$_{\gamma}$/F$_\mathrm{X}$ in this way is that the flux of the Chandra blob (extraction radius of $\sim$50'') is compared to the VHE flux of a more extended object ($\theta \sim 0.15^\circ$).

\begin{table*}
\begin{center}
\begin{tabular}{|ccccccc|}

\hline
src	& CXOU name		& RA (J2000)	& Dec(J2000)	& countrate			& possible identification	& 2MASS sources	\\
\hline
1	& CXOU J150850.6-621018	& 15:08:50.637 & -62:10:18.24	& $(4.6 \pm 0.2)\times10^{-2}$	& 1RXS J150841.2-621006		& \\
\hline
2	& CXOU J150752.0-621509	& 15:07:52.091 & -62:15:09.96	& $(1.0 \pm 0.3)\times10^{-3}$	& 3UC 056-316462		& 15075219-6215107\\
	& 	&  & 	& 									& GSC2.3 S7QP004587 		& 15075188-6215057\\
	& 	&  & 	& 									& GSC2.3 S7QP084980		& 15075290-6215077\\
\hline
3	& CXOU J150718.5-621357	& 15:07:18.516 & -62:13:57.22	& $(1.4 \pm 0.3)\times10^{-3}$	& 3UC 056-316221		& 15071853-6213571\\
	& 	&  & 	& 									& GSC2.3 S7QP005118 		& \\
\hline
4	& CXOU J150621.7-621110	& 15:06:21.789 & -62:11:10.76	& $(1.7 \pm 0.4)\times10^{-3}$	& TYC 9024.1615.1		& 15062140-6211110\\
	& 	&  & 	& 									& 3UC 056-315763 		& \\
\hline
5	& CXOU J150636.9-621628	& 15:06:36.935 & -62:16:28.47	& $(1.0 \pm 0.3)\times10^{-3}$	& 	-			&  -\\
\hline
6	& CXOU J150645.7-621648	& 15:06:45.788 & -62:16:48.10	& $(2.5 \pm 0.4)\times10^{-3}$	& GSC2.3 S7QP084624		& 15064587-6216466\\
	& 	&  & 	& 									& GSC2.3 S7QP084624		& \\
\hline
7	& CXOU J150708.8-621643	& 15:07:08.824 & -62:16:43.99	& $(7.6 \pm 0.8)\times10^{-3}$	& TYC 9024.1705.1		& 15070879-6216441\\
	& 	&  & 	& 									& 3UC 056-316134	 	& \\
\hline
8	& CXOU J150706.7-621858	& 15:07:06.746 & -62:18:58.75	& $(2.5 \pm 0.4)\times10^{-3}$	& GSC2.3 S7QP060692		& 15070693-6218560\\
\hline
9	& CXOU J150756.0-622238	& 15:07:56.018 & -62:22:38.70	& $(4.2 \pm 0.7)\times10^{-3}$	& GSC2.3 S7QP035739		& 15075554-6222336\\
	& 	&  & 	& 									& GSC2.3 S7QP035813 		& 15075683-6222411\\
\hline
10	& CXOU J150656.1-623040	& 15:06:56.128 & -62:30:40.72	& $(1.7 \pm 0.4)\times10^{-3}$	& GSC2.3 S7QP031378		& 15065496-6230425\\
	& 	&  & 	& 									& GSC2.3 S7QP031375 		& 15065736-6230444\\
	& 	&  & 	& 									& GSC2.3 S7QP031476 		& 15065721-6230287\\
	& 	&  & 	& 									& GSC2.3 S7QP001419 		& 15065613-6230553\\
	& 	&  & 	& 									& GSC2.3 S7QP031426 		& 15065368-6230417\\
	& 	&  & 	& 									& GSC2.3 S7QP031294 		& 15065363-6230347\\
	& 	&  & 	& 									& 		 		& 15065746-6230579\\
\hline
11	& CXOU J150606.7-622210	& 15:06:06.799 & -62:22:10.36	& $(1.7 \pm 0.4)\times10^{-3}$	& 	-			& 15060774-6222092\\
\hline
12	& CXOU J150538.1-622001	& 15:05:38.114 & -62:20:01.35	& $(1.3 \pm 0.3)\times10^{-3}$	& GSC2.3 S7QP003678		& 15053757-6220015\\
\hline

\end{tabular}
\end{center}
\label{gmbperf}
\caption{Summary of the twelve surces detected by Chandra using {\tt celldetect}. For source 1 only the firm identification 1RXS J150841.2-621006 (\cite{faintRASS}) is mentioned. The Tycho-2 catalogue (\cite{tycho2}), the second-generation guide star catalog (\cite{gsc}), the UCAC3 Catalogue (\cite{ucac3}), and the 2MASS All-Sky Catalog of Point Sources (\cite{2mass}) have been inspected when searching for possible counterparts.  }
\label{gmbperf22}
\end{table*}

If the previously discussed association of the faint extended X-ray source and HESS J1507-622 cannot be maintained, e.g. as the result of deeper or complementary multifrequency observations, we could also be faced with the case of not having any counterpart at all. In this case, HESS J1507-622, according to its position and to its brightness in VHE gamma-rays, could be a candidate for a ``dark accelerator'', i.e. of a purely hadronic source, visible mainly at TeV energies.
On the other hand, it will also be argued that a mature PWN can appear to look like a ``dark accelerator''; in this case, the emission would be from inverse Compton scattering, coming from a TeV calorimeter around the aging pulsar (\citep[see e.g.][]{d08}).

\subsection{Leptonic scenario: PWN}

 In a leptonic scenario it is expected that the same population of high-energy electrons that generate the gamma-ray emission should also produce nonthermal X-rays due to synchrotron radiation caused by the ambient magnetic field. The non-detection of X-ray emission from the location of HESS J1507-622 would therefore place limits on the strength of any magnetic field at the source. If we use the X-ray brightness of 1RXS J150841.2-621006 to place an upper limit on any synchrotron component connected to the detected VHE emission and if we assume that the gamma-ray emission is produced by electrons following a power-law distribution with a spectral index of -2.2 that up-scatter cosmic microwave background (CMB) photons, then the ambient magnetic field cannot exceed a strength of 0.5 $\mu$G. The caveat mentioned in the previous section applies here since we have determined the magnetic field by comparing the flux of a ROSAT point source with the VHE flux of an extended object. For higher upper limits for any diffuse X-ray emission, the limit on the magnetic field is less constraining.

PWNe are usually considered natural candidates for leptonic Galactic gamma-ray emitters (\citep[e.g.][]{gaensler06}). Although they are, in many cases, detected as nonthermal X-ray sources, in the following we outline how an evolved PWN can indeed lead to a fairly bright gamma-ray source without any counterpart (\cite{d08}). The key issue
for such a model is that the low energy synchrotron emission, where $\tau_E$ is the synchrotron emitting lifetime of TeV gamma-ray emitting leptons with energy E, depends on an internal property of the PWN, namely the magnetic field (\citep[e.g.][]{kennel}) which may vary as a function of time, following $\tau_E\propto t^{2\alpha}\;\;{\rm if}\;\;B(t)\propto t^{-\alpha},$ where $\alpha$ is the power-law index of the decay of the average nebular field strength, whereas the VHE emission depends on the CMB radiation field, which is constant on timescales relevant for PWN evolution.
Magnetohydrodynamical (MHD) simulations of composite SNRs with different spindown power at birth and spindown timescale (\cite{fd09}) find that $\alpha=1.3$ until the passage of the reverse shock (at time $t=T_R$), when $B(t)$ flattens out, or even increases slightly. As the simulation time only extended up to 10 kyr (with $B(t)\sim B(T_R)$), it is expected that field decay would continue for $t\gg T_R$ since expansion continues even after the passage of the reverse shock. 

From hydrodynamic simulations, an expression for the time of the reverse shock passage (i.e. the return time of the reverse shock to the origin) has been given (\cite{fd08}) as
\begin{eqnarray*}\label{vergelyking}
T_{R}=10\;{\rm kyr} \left( \frac{\rho_{ISM}}{10^{-24} {\rm g/cm}^{-3}}\right)
^{-1/3} \left( \frac{M_{ej}}{10M_{\odot}}\right) ^{3/4}  \left(
\frac{E_{ej}}{10^{51}{\rm erg}}\right) ^{-2/3}
\end{eqnarray*}
where $\rho_{ISM}$ is the density of the interstellar medium, $E_{ej}$ the SNR blast wave energy and $M_{ej}$ the ejecta mass during the SNR explosion. The stellar wind of a high-mass star can blow a cavity around the progenitor star (\cite{bubble}) with relatively low ISM density, so that $T_R\gg 10$ kyr. In such a case it is expected that $B(t)$ can decay as $t^{-\alpha}$, until the field is low enough for the X-ray flux to drop below the typical sensitivity levels, and that $t\gg \tau_0$ (where $\tau_0$ is the spindown timescale), while $t<T_R$, so that the time integral of the total measured energy of nebular leptons is convergent; i.e., the initial kinetic energy at birth of the pulsar is measured. As a result, in a scenario where the magnetic field decays as a function of time, the synchrotron emission will also fade as the PWN evolves. 
The reduced synchrotron losses for high-energy electrons for such a scenario will then lead to increased lifetimes for these leptonic particles. For timescales shorter than the inverse-Compton lifetime of the electrons (t$_{IC}$ $\approx$ 1.2 $\times$ 10$^6$ (E$_e$/1 TeV)$^{-1}$ years), this will result in an accumulation of VHE electrons which will also lead to an increased gamma-ray production due to up-scattering of CMB photons. Such accumulation of very-high energy electrons in a PWN has indeed been seen in the source HESS J1825-137 (\cite{aharonian06}). To summarize, during their evolution PWN may appear as gamma-ray sources with only very faint low-energy counterparts and this may represent a viable model for HESS J1507-622.

Within this scenario, the faint diffuse emission detected by Chandra assumes a particular importance. The morphology of HESS J1507-622 is similar to HESS J1702-420 (\cite{unid}), where the pulsar PSR J1702-4128 could be powering an extremely asymmetric PWN visible in VHE gamma-rays, but it should be emphasized that the PWN association of HESS J1702-420 is not established. In the ancient PWN scenario therefore the faint diffuse emission in X-ray could be due to a younger population of electrons and reflect the proper motion of the pulsar within the nebula. Also the F$_{\gamma}$(1-10 TeV)/F$_\mathrm{X}$(2-10 keV) is similar in these two VHE unidentified sources; in fact, \cite{fuji} calculated F$_{\gamma}$(1-10 TeV)/F$_\mathrm{X}$(2-10 keV)$> 32$ for HESS J1702-420. 


In the framework of a PWN interpretation, it is also interesting to compare HESS J1507-622 with other VHE PWNe and PWN candidates that are located at a similar angular offset from the Galactic plane. These objects are (sorted by decreasing offset) Crab (b = -5.8$^\circ$) (\cite{weekes89}), Geminga (b = 4.3$^\circ$) (\cite{abdo07}), Vela X (b = -3.1$^\circ$) (\cite{velax}) and HESS J1356-645 (b = -2.5$^\circ$) (\cite{renaud08}). All these objects related to prominent pulsars and to their associated multiwavelength counterparts. Two of them are nearby sources (Geminga with d = 160 pc, \cite{caraveo96}, and Vela with d = 290 pc, \cite{caraveo01}) and feature very extended VHE emission regions (Geminga $\sim 2.8^\circ$, as indicated in \cite{abdo07}, and Vela X $\sim 0.5^\circ$ radius, as measured by H.E.S.S.), as can be expected for close, extended objects. Geminga is the only example here of an old pulsar: its age has been estimated to be $\sim$ 300 000 years (\cite{bertsch92}). In comparison to all these sources, HESS J1507-622 is different since it lacks any obvious counterpart. If it is indeed an evolved PWN, then its rather compact appearance would disfavor a very small distance; by simple comparison with Geminga, we should conclude that the PWN scenario constrains HESS J1507-622 distance to a value $>$6 kpc.

\subsection{Hadronic scenario: SNR and GRB remnant}


For the hadronic scenario of gamma-ray production, the total energy in cosmic rays
in the object can be derived from the gamma-ray luminosity and the density
of target material. Since HESS J1507-622 is located off the Galactic disk, the
density of target material is lower than in the Galactic plane and can be obtained
as a function of distance, following the density profile of the interstellar medium perpendicular to
the Galactic plane. Using the best-fitting model (\cite{lockman84}) for the density profile off the Galactic plane at
relevant galactocentric distances between 4 and 8 kpc and assuming a
spectral index of -2 and a cut-off at 100 TeV for the cosmic rays, it is
found that about 10$^{50}$ ($\sim10^{51}$) erg of cosmic rays at a distance of
2 (4) kpc are needed to explain the VHE emission observed in HESS J1507-622.
The very fast increase in the energy in cosmic rays with distance (a
factor of 10 for an increase in distance by a factor of 2) results from
the fact that a larger distance places the source also at a larger
spatial offset from the Galactic disk, hence in an environment with
lower density of target material.

The main accelerators of hadronic cosmic rays are believed to be the shock waves
of SNRs. If HESS J1507-622 is indeed connected to an SNR, due to its energetics, it
would be located at a distance of $\lesssim$ 2 kpc  (density of 0.45 cm$^{-3}$) (\cite{lockman84}). A supernova remnant at a
distance of  2 kpc with the apparent angular size of HESS J1507-622
would be in the pressure-driven Sedov phase with an age of less than 1000
years. It is interesting to note that such an SNR should
feature similar observable properties as the remnant of SN 1006. However,
SN 1006 is a prominent emitter of nonthermal X-ray emission, e.g. ASCA (\cite{koyama96}) and ROSAT (\cite{rosat1006}), and the absence of such X-ray emission in HESS J1507-622
disfavors any SNR origin for this VHE source. In this scenario the faint X-ray extended emission could be part of an SNR shell.

A proposed more exotic class of hadronic VHE gamma-ray sources are
remnants of gamma-ray bursts (\cite{atoyan06}). If HESS
J1507-622 is indeed the result of a GRB, it would be of the subclass of
compact binary merger induced short GRBs since the location of HESS J1507-622
off the Galactic plane would indicate that it is connected to an old
stellar population (\cite{domainko05}, \cite{domainko08}). Also the X-ray afterglows of short bursts indicate a low-density environment that would be consistent with a location away from galactic disks (see e.g. \cite{nakar07}). In the model of a GRB remnant, hadronic cosmic rays
are produced in a point-like explosion and extended, center-filled VHE
gamma-ray sources are then the result of energy dependent diffusion of the
cosmic rays and subsequent hadronic gamma-ray production. In this picture
the age of the remnant can be estimated from the source extension to
$t/1 ~\mathrm{kyr} \lesssim 0.2 (d/1 ~\mathrm{kpc})^2$ assuming typical values of
Galactic cosmic ray diffusion (\cite{atoyan06}), which may be different outside the Galactic plane. The rate of merger-induced bursts in the Galaxy is found to be about one event every (0.5 - 7) $\times$ 10$^4$ years (\cite{kalogera04}, \cite{guetta06}). For an anticipated age of $\gtrsim 10^4$ years the remnant has
to be at a distance of $\gtrsim$ 7 kpc. At such a distance the total
energy of cosmic rays would be $\sim 10^{52}$ erg, which is quite high for
the energetics of typical short bursts (\cite{nakar07}). To summarize, a GRB remnant interpretation of HESS J1507-622
would require a highly energetic and relatively recent merger event ($\lesssim 10^4$ years) which
may be possible, but is quite unlikely.

\section{Conclusions}

The discovery of HESS J1507-622, a rather bright ($\sim$8~\% of the Crab flux) unidentified Galactic VHE gamma-ray source, is reported. The lack of obvious counterparts, together with its location 3.5$^{\circ}$ off the Galactic plane, play a crucial role in interpretating this source, which shows the highest value yet found for the ratio between X-ray and VHE emission. Hadronic and leptonic scenarios for the VHE gamma-ray radiation are discussed, showing that a PWN scenario with an old pulsar could be a possible interpretation for HESS J1507-622, while a hadronic  model (e.g., SNR-related hadronic emission and GRB remnant) appears unlikely unless the distance is fairly small. 
Upcoming deeper X-ray observations (XMM-Newton and Suzaku) will undoubtedly offer deeper insight in this VHE source.


\begin{acknowledgements}

The support of the Namibian authorities and of the University of Namibia in facilitating the construction and operation of HESS is gratefully acknowledged, as is the support by the German Ministry  for Education and Research (BMBF), the Max Planck Society, the French Ministry for Research, the CNRS-IN2P3, and the Astroparticle Interdisciplinary Programme of the CNRS, the U.K. Science and Technology Facilities Council (STFC), the IPNP of the Charles University, the Polish Ministry of Science and Higher Education, the South African Department of Science and Technology and National Research Foundation, and by the University of Namibia. We appreciate the excellent work of the technical support staff in Berlin, Durham, Hamburg, Heidelberg, Palaiseau, Paris, Saclay, and in Namibia in the construction and operation of the equipment. Finally O. T. would like to acknowledge P. Caraveo for her constructive feedback in the referee process, as well as P. Slane, J. Halpern and M. Roberts for the very useful discussions.

\end{acknowledgements}


\begin{thebibliography}{}

   \bibitem[Abdo et al. 2007]{abdo07} Abdo, A. et al. (MILAGRO Collaboration) 2007, ApJ, 664, L91

   \bibitem[Aharonian et al. 2004]{calibration} Aharonian, F. et al. (H.E.S.S. Collaboration) 2004,
      Astropart. Phys., 22, 109

   \bibitem[Aharonian et al. 2005]{survey} Aharonian, F. et al. (H.E.S.S. Collaboration) 2005,
      Science, 307, 1839

\bibitem[Aharonian et al. 2006a]{survey2} Aharonian, F. et al. (H.E.S.S. Collaboration)
2006, ApJ, 636, 777

\bibitem[Aharonian et al. 2006b]{velax} Aharonian, F. et al. (H.E.S.S. Collaboration)
2006, A\&A, 448, L43

   \bibitem[Aharonian et al. 2006c]{crab} Aharonian, F. et al. (H.E.S.S. Collaboration) 2006,
      A\&A, 457, 899

   \bibitem[Aharonian et al. 2006d]{aharonian06} Aharonian, F. et al. (H.E.S.S. Collaboration) 2006, A\&A, 460, 365

   \bibitem[Aharonian et al. 2008]{unid} Aharonian, F. et al. (H.E.S.S. Collaboration) 2008,
      A\&A, 477, 353


   \bibitem[Atoyan et al. 2006]{atoyan06} Atoyan, A., Buckley, J., and Krawcynski, H. 2006,
ApJ, 642, L153

  \bibitem[Benjamin 2005]{Spitzer} Benjamin, R. A. 2005,
AAS, 207.6307B

   \bibitem[Berge et al. 2007]{berge} Berge, D., Funk, S., and Hinton, J. 2007,
       A\&A, 466, 1219

   \bibitem[Bertsch et al. 1992]{bertsch92} Bertsch, D. L., Brazier, K. T. S., Fichtel, C. E. et al. 1992, Nature, 357, 306

   \bibitem[Caraveo et al. 1996]{caraveo96} Caraveo, P. A., Bignami, G. F., Mignani, R., Taff, L. G. 1996, ApJ, 461, L91

   \bibitem[Caraveo et al. 2001]{caraveo01} Caraveo, P. A., De Luca, A.,  Mignani, R. P., Bignami, G. F.  2001, ApJ, 561, 930


   \bibitem[Chevalier \& Liang 1989]{bubble} Chevalier, R.A., and Liang, P. 1989,
 ApJ, 344, 332

   \bibitem[Cutri et al. 2003]{2mass} Cutri, R. M. et al. 2003, 2MASS All Sky Catalog of point sources

   \bibitem[Dame et al. 2001]{Dame} Dame, T. M., Hartmann, Dap, and Thaddeus, P. 2001,
      Apj, 547, 792

   \bibitem[de Jager 2008]{d08} de Jager, O.C. 2008, ApJ, 678, L113

\bibitem[de Jager et al. 2009]{fd09} de Jager, O.C. et al. 2009, in the $31^{st}$ International Cosmic Ray Conference proceedings, arXiv:astro-ph/0906.2644

   \bibitem[de Naurois 2006]{model2D} de Naurois, M. 2006,
        arXiv:astro-ph/0607247

   \bibitem[Domainko \& Ruffert 2005]{domainko05} Domainko, W., and Ruffert, M. 2005, A\&A, 444, L33

   \bibitem[Domainko \& Ruffert 2008]{domainko08} Domainko, W., and Ruffert, M. 2008, AdSpR, 41, 518

   \bibitem[Duncan et al. 1995]{duncan} Duncan, A. R., Stewart, R. T., Haynes, R. F. and Jones, K. L. 1995, MNRAS, 277, 36

   \bibitem[Ferreira \& de Jager 2008]{fd08} Ferreira, S.E.S., and de Jager, O.C. 2008,
 A\&A, 478, 17


   \bibitem[Fujinaga et al. 2009]{fuji} Fujinaga T. et al. 2009, in ``The Energetic Cosmos: from Suzaku to ASTRO-H'' conference.


   \bibitem[Gaensler \& Slane 2006]{gaensler06} Gaensler, B. M. and Slane, P. O. 2006, ARA\&A, 44, 17

   \bibitem[Green et al. 1999]{molonglo} Green, A.J. et al. 1999, ApJS, 122, 207

   \bibitem[Guetta \& Piran 2006]{guetta06} Guetta, D. and Piran T. 2006, A\&A, 453, 823


   \bibitem[Harris et al. 1998]{ROSATsys} Harris D. E. et al. 1998, A\&AS, 133, 431

   \bibitem[Haverkorn et al. 2006]{haverkorn} Haverkorn, M et al. 2006, ApJS, 167, 230

   \bibitem[Hog et al. 2000]{tycho2} Hog, E. et al. 2000, A\&A, 335, L27

   \bibitem[Kalogera et al. 2004]{kalogera04} Kalogera, V., Kim, C., Lorimer, D. R., et al.
2004, ApJ, 601, L179, erratum: 2004 ApJ, 614, L137

   \bibitem[Kargaltsev \& Pavlov 2010]{kar} Kargaltsev, O. and Pavlov, P. O. 2010, AIP Conference Series, 1248, 25

   \bibitem[Kennel \& Coroniti]{kennel} Kennel, C. F. and Coroniti, F. V. 1984, ApJ, 283, 710.

   \bibitem[Koyama et al. 1996]{koyama96} Koyama, K. et al. 1996, Nature, 378, 255

   \bibitem[Lasker et al. 2008]{gsc} Lasker, B. M. 2008,
      AJ, 136, 735

   \bibitem[Li \& Ma 1983]{LiMa} Li, T., and Ma, Y. 1983,
      ApJ, 272, 317

   \bibitem[Lockman 1984]{lockman84} Lockman, F. J. 1984, ApJ, 283, 90

\bibitem[2007]{matsumoto07} Matsumoto, H., Ueno, M., Bamba, A. et al. 2007, PASJ, 59, S199

   \bibitem[Nakar 2007]{nakar07} Nakar, E. 2007, PhR, 442, 166


\bibitem[Renaud et al. 2008]{renaud08} Renaud, M. et al. (H.E.S.S. Collaboration) 2008, AIPC, 1085, 285


    \bibitem[Simon et al. 2006]{msx} Simon, T. et al. 2006
      Apj, 639, 227





   \bibitem[Voges et al. 2000]{faintRASS} Voges, W. et al. 2000,
      IAU Circ., 7432, 1

   \bibitem[Weekes et al. 1989]{weekes89} Weekes, T. C. et al. (Whipple Collaboration)
1989, ApJ, 342, 379

   \bibitem[Willingale et al. 1996]{rosat1006} Willingale, R. et al. 1996 
      MNRAS, 278, 749

\bibitem[2006]{yamazaki06} Yamazaki, R., Kohri, K., Bamba, A. et al. 2006, MNRAS, 371, 1975

   \bibitem[Zacharias et al. 2008]{ucac3} Zacharias, N. et al. 2008, AJ, UCAC3 Catalogue






\end{thebibliography}
\end{document}